\documentclass[letterpaper, 10 pt, conference]{ieeeconf}  

\IEEEoverridecommandlockouts                              

\usepackage{xcolor}

\usepackage{cite}
\usepackage{multicol}
\usepackage{amsmath,amssymb,amsfonts}
\usepackage{algorithmic}
\usepackage{eqnarray,amsmath}
\usepackage{xcolor}
\usepackage{graphicx}
\usepackage{siunitx}
\usepackage{textcomp}
\usepackage{makecell}
\usepackage{hyperref}
\usepackage{multirow}
\usepackage{tabularx}

\usepackage{array}
\usepackage{kantlipsum}

\usepackage{array}

\makeatletter
\newcommand{\thickhline}{%
    \noalign {\ifnum 0=`}\fi \hrule height 1pt
    \futurelet \reserved@a \@xhline
}
\makeatother

\title{\LARGE \bf
Prototype-based Domain Generalization Framework\\
for Subject-Independent Brain-Computer Interfaces
}

\author{Serkan Musellim$^{1}$, Dong-Kyun Han$^{1}$, Ji-Hoon Jeong$^{2}$, and Seong-Whan Lee$^{3}$, 
\IEEEmembership{Fellow, IEEE}
\thanks{\copyright 20XX IEEE.  Personal use of this material is permitted. Permission from IEEE must be obtained for all other uses, in any current or future media, including reprinting/republishing this material for advertising or promotional purposes, creating new collective works, for resale or redistribution to servers or lists, or reuse of any copyrighted component of this work in other works.}
\thanks{This work was partly supported by Institute of Information and Communications Technology Planning and Evaluation (IITP) grants funded by the Korea government (No. 2015-0-00185, Development of Intelligent Pattern Recognition Softwares for Ambulatory Brain-Computer Interface; No. 2017-0-00451, Development of BCI based Brain and Cognitive Computing Technology for Recognizing User’s Intentions using Deep Learning;  No. 2021-0-00866, Development of BMI application technology based on multiple bio-signals for autonomous vehicle drivers).}
\thanks{$^{1}$Serkan Musellim and Dong-Kyun Han are with the Department of Brain and 
Cognitive Engineering, Korea University, Republic of Korea. 
        {\tt\small \{serkanmusellim, dk\_han\}@korea.ac.kr}}%
\thanks{$^{2}$Ji-Hoon Jeong is with the Department of Computer Science, Chungbuk National University, Republic of Korea. 
{\tt\small jh.jeong@chungbuk.ac.kr}}
\thanks{$^{3}$Seong-Whan Lee is with the Department of Artificial Intelligence, Korea 
University, Republic of Korea. 
{\tt\small sw.lee@korea.ac.kr}}
}

\begin{document}

\maketitle
\thispagestyle{empty}
\pagestyle{empty}

\begin{abstract}

Brain-computer interface (BCI) is challenging to use in practice due to the inter/intra-subject variability of electroencephalography (EEG). The BCI system, in general, necessitates a calibration technique to obtain subject/session-specific data in order to tune the model each time the system is utilized. This issue is acknowledged as a key hindrance to BCI, and a new strategy based on domain generalization has recently evolved to address it.
In light of this, we've concentrated on developing an EEG classification framework that can be applied directly to data from unknown domains (i.e. subjects), using only data acquired from separate subjects previously.
For this purpose, in this paper, we proposed a framework that employs the open-set recognition technique as an auxiliary task to learn subject-specific style features from the source dataset while helping the shared feature extractor with mapping the features of the unseen target dataset as a new unseen domain. Our aim is to impose cross-instance style in-variance in the same domain and reduce the open space risk on the potential unseen subject in order to improve the generalization ability of the shared feature extractor. 
Our experiments showed that using the domain information as an auxiliary network increases the generalization performance. 

\indent \textit{Clinical relevance}—This study suggests a strategy to improve the performance of the subject-independent BCI systems. Our framework can help to reduce the need for further calibration and can be utilized for a range of mental state monitoring tasks (e.g. neurofeedback, identification of epileptic seizures, and sleep disorders).

\end{abstract}

\section{INTRODUCTION}

Brain-computer interface (BCI) is a platform for communicating with a computer system utilizing human brain impulses without the need for muscles. Because of its affordable and portable nature,  electroencephalogram (EEG) signals have been widely used in many clinical and research BCI applications \cite{1}. 
By using specifically developed signal processing and machine learning algorithms, the collected waveforms are carefully integrated and collectively utilized to produce a real-time control signal \cite{mason2007comprehensive} for further BCI processing.  

EEG analysis is complicated since variations in the user's physiological or psychological state cause fluctuations over time, and it differs from subject to subject. The model's performance on unknown data is hampered by the inter/intra-subject variability of EEG signals.
As a result, generic BCI systems need a calibration technique to collect subject/session-specific data in order to fine-tune the model each time the device is utilized \cite{lotte2010regularizing,jeong2020decoding,discrim2016}. EEG components such as rapid serial visual presentation, steady-state visual evoked potentials, and event-related potentials require fewer calibration data or do not require any calibration at all \cite{rsvp,erp}. On the other hand, motor imagery (MI) and speech imagery need well-developed calibration methods. Although MI is a practical modality for patients with motor disabilities since it requires no muscular activities, the calibration process of MI might take up to an hour of data collecting and model training time, which is unpleasant for the new user \cite{motorimagery,robotarm}. 

For this purpose, there have been transfer learning-based approaches that use other subjects' data to help learn a small amount of calibration data or to create a generalized model\cite{multisubjectEMBC, ensembleCNN}.
Considering the variability problem from the perspective of the domain shift problem, it is divided into domain adaptation and domain generalization depending on whether the data of the target subject is used.
This subject-independent BCI corresponds to domain generalization because calibration data from the target subject is not used at all during the entire training procedure.
Most of the studies in this area use the leave-one-subject-out validation, which uses data of all users other than the target in order to use as much data as possible.
However, we cannot rule out the possibility that subjects that may have a negative effect are included in the transfer learning process that uses the data of others. This may even happen with the data of the same subjects if the data is collected on different days/sessions. For this reason, it may be beneficial for the network to recognize the samples that belong to the same subject and make their features as close as possible.

To deal with that problem, the invariance cross instances from the same subject can be targeted, and, according to our hypothesis, the feature extractor can be trained better for an unseen subject with an open-set recognition (OSR) technique. There is a possibility of encountering an unknown task or object in real-world circumstances. The goal of OSR is to provide a system that can correctly identify unknown and known classes during testing. The main goal is to simultaneously lower the empirical classification risk on labeled known data and the open space risk on potentially unknown data\cite{towardosp}. In traditional deep-learning methods, a linear classification layer is applied to the features by applying the softmax function on the obtained features. This operation divides the feature space with hyperplanes into subspaces. The number of subspaces is equal to the number of classes. Although the softmax operation separates the classes well, it is not good at telling the difference between the known and unknown classes. For this reason, several methods\cite{cpl, cpn} use prototype learning to force the training features to be close to corresponding prototypes, which makes it easier to distinguish the known and unknown classes.

In this study, we proposed a prototype-based domain generalization framework by using an auxiliary network that employs the OSR technique to learn subject-specific style features from the source dataset while helping the feature extractor with mapping the features of the unseen target dataset as a new unseen domain.

\section{METHODS}

\begin{figure}[t!]
    \centering
    \includegraphics[width=0.9\linewidth]{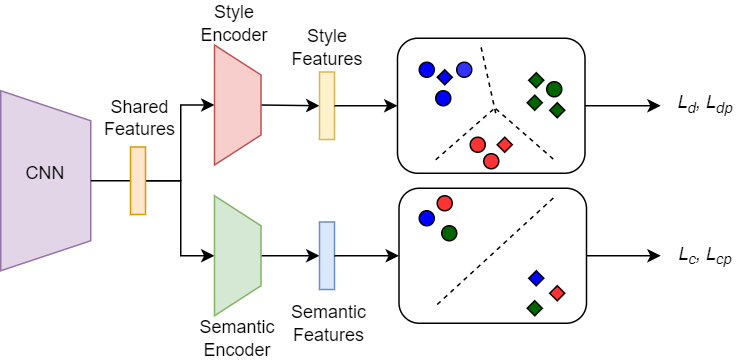}
    \caption{Framework of the proposed method. The style encoder extracts the subject information, the semantic encoder extracts the class/task information. In feature space different colors refer to different subjects, different shapes refer to different classes.}
    \label{fig1:Framework}
\end{figure}

Rather than using the subject information in an adversarial manner, our idea is to impose cross-instance style in-variance in the same domain by utilizing the subject labels. For this purpose, a style encoder follows the convolutional neural network (CNN) along with the semantic encoder. CNN acts as a shared featured extractor between these encoders. The purpose of the style encoder is to help the feature extractor recognize the subject and support the classification. In order to reduce the open space risk on the potential unknown subject and to have a better generalized network, convolutional prototype learning (CPL) \cite{cpl} is adapted as the OSR method to classify the subjects. In addition, we utilize CPL for task classification on the semantic features that come from the semantic encoder as well. The whole framework can be seen in Fig. \ref{fig1:Framework}. It should be highlighted that our goal is not to classify the subjects perfectly, but rather to assist with task classification using knowledge coming from the subjects.

\subsection{Convolutional Prototype Learning for EEG Decoding}

Originally, CPL is a framework designed to handle the open-world problem and improve the robustness of the feature extractor by utilizing prototypes. Different from the traditional CNN, in CPL, the softmax layer is removed, and prototypes are learned from the data. These prototypes are jointly trained with the feature extractor and instances are classified into the class with the nearest prototype. As the distance metric between the prototypes and instances, we used Euclidean distance in this study just as in the original paper.

While training the feature extractor we used distance-based cross-entropy loss (DCE)\cite{cpl}. In this approach, the distance between the samples and the prototypes is treated as the probability of that sample belonging to the class of the corresponding prototype. Let's say we have $C$ number of classes and have one prototype for each class. For a training sample $(x,y)$, let $m_{i}$ denote the prototype where $i$ $\in$ ($1,2,...,C$) and $M = \{m_i | i=1,...,C\}$ be the parameters of the prototypes that are learned during training. Feature extractor is denoted as $f(x,\theta)$, where $\theta$ is the parameters of the network. In our case; CNN and style encoder are the feature extractor for style features and CNN and semantic encoder are the feature extractor for class/task features. Then the probability in proportion can be written as:

\begin{equation}
p(x \in m_i\mid x) \propto (\parallel f(x) - m_i \parallel)^2_2
.\end{equation}
Furthermore, in order to satisfy the sum-to-one property of probability, the distances are passed to a softmax equation as the following:

\begin{equation}
p(y | x) = \frac{e^{-\gamma(\parallel f(x) - m_i \parallel_2^2)}}{\sum_{k=1}^C e^{-\gamma(\parallel f(x) - m_k \parallel_2^2)}}
.\end{equation}
Here $\gamma$ is the temperature hyper-parameter that controls the hardness of the softmax operation. Based on this loss function cross-entropy loss is defined.

Lastly, a prototype loss (PL) is added to the equation as a regularizer to improve the generalization performance of the model. This loss function is defined as:

\begin{equation}
l_p((x,y);\theta, M ) = \parallel f(x) - m_i \parallel_2^2
 .\end{equation}
The total loss function can be summarized as:

\begin{equation}
loss((x,y);\theta, M ) = l((x,y);\theta, M ) + \beta l_p((x,y);\theta, M ),
\end{equation}
where $\beta$ is the hyper parameter that controls the weight of the prototype loss.

\subsection{Open-Set Recognition for Subject Identification}

In general, subject information is ignored or used in an adversarial manner in subject-independent EEG networks. In this study, we try to get the advantage of this information as an auxiliary network to support the classification of the main task. We assume that by learning the subject information CNN will be better trained for that subject's data. To be able to use this method on an unseen subject, we propose to use a classification technique that is used in the open-set recognition area. In our hypothesis, by using an OSR method on the auxiliary network CNN will be better trained for the unseen subject.
Therefore, CPL has been used as an OSR method to classify the subjects.

In order to have the subject information separately from the class information, we utilize two encoders called style and semantic encoders that take the output of CNN. The output of the style encoder is tried to match with centers that are learned from the subject labels and the output from the semantic encoder is tried to match with centers that are learned from the task labels (i.e. motor imagery classes). The total loss function of our framework can be summarized as below:

\begin{equation}
\mathcal{L} = \mathcal{L}_c + \beta_1 \mathcal{L}_{cp} + \alpha( \mathcal{L}_d + \beta_2 \mathcal{L}_{dp}),
\end{equation}
where $\mathcal{L}_c$ is the DCE loss for the task, $\mathcal{L}_{cp}$ is the prototype loss for the task, $\mathcal{L}_d$ is the DCE loss for subjects, $\mathcal{L}_{dp}$ is the prototype loss for subjects, and $\alpha$ is the hyper-parameter that controls the weight of the subject loss.

\section{EXPERIMENTS}

\subsection{Dataset} 

OpenBMI dataset \cite{openbmi}: This dataset consists of 54 subjects and 2 motor imagery classes (left hand, right hand). Each subject has 4 sessions of data, each of these sessions include 100 trials (200 trials for a class). Each trial consists of 4 seconds of data that is sampled with 1,000 Hz and collected from 62 electrodes. We use 4 seconds of task data and all 62 channels. The signals are downsampled to 250 Hz from 1,000 Hz with an order-8 Chebyshev type-I filter for anti-aliasing.

\subsection{Data-Split}

We validated our experiments by leaving one subject out.
For example, the source data for cross-subject training was the samples excluding one subject from the 54 subjects' samples (i.e. 53 subjects), and the excluded data was the test data.
We divided the source dataset into two parts, each with an 80:20 split, and used them as train and validation data, respectively.
Only the $4^{th}$ session of the test subject was used to evaluate the network as the other research\cite{adaptivenn}. The rest of the test data was kept aside for a possible future adaptation scenario. We evaluated the proposed method and the baseline methods on 4 different sizes of training data. Namely, 11, 21, 31, and 54 subjects. In all settings, the test subject was excluded and the rest was used for train and validation. We have 10, 20, 30, and 53 subjects for training respectively. These subjects were chosen randomly but all the frameworks were trained with the same set of subjects.

\begin{table}[t]
\caption{mi classification results for subject-dependent and subject-independent methods (statistical significance
($p < 0.05$) between the proposed framework and other frameworks are marked with *)}
\label{tab:results}
\resizebox{\columnwidth}{!}{%
\begin{tabular}{l|cccc}
\hline
\multicolumn{1}{c|}{\textbf{Methods}}                                            & \multicolumn{4}{c}{\textbf{Accuracy (\%) (Standard Deviation)}}                                                                                                                                                                                                                                                                                                                                                        \\
\thickhline\textbf{Subject Dependent}   & \multicolumn{1}{l}{}                                                                                  & \multicolumn{1}{l}{}                                                                                  & \multicolumn{1}{l}{}                                                                                  & \multicolumn{1}{l}{}                                                              \\
\thickhline
CSP\cite{csp}                & \multicolumn{4}{c}{68.17 ($\pm$17.57)}                                                                                                                                                                                                                                                                                                                                                                             \\ \hline
CSSP\cite{cssp}                                        & \multicolumn{4}{c}{69.69 ($\pm$18.53)}                                                                                                                                                                                                                                                                                                                                                                             \\ \hline
FBCSP\cite{fbcsp}                                      & \multicolumn{4}{c}{70.59 ($\pm$18.56)}                                                                                                                                                                                                                                                                                                                                                                             \\ \hline
BSSFO\cite{bssfo}                                      & \multicolumn{4}{c}{71.02 ($\pm$18.83)}                                                                                                                                                                                                                                                                                                                                                                             \\ \hline
CNN\cite{freqcnn}                                      & \multicolumn{4}{c}{71.32 ($\pm$15.88)}                                                                                                                                                                                                                                                                                                                                                                             \\ \hline
DeepConvNet\cite{adaptivenn}                           & \multicolumn{4}{c}{63.54 ($\pm$14.25)}                                                                                                                                                                                                                                                                                                                                                                             \\
\thickhline\textbf{Subject Independent} & \multicolumn{1}{c}{}                                                                                  & \multicolumn{1}{c}{}                                                                                  & \multicolumn{1}{c}{}                                                                                  &                                                                                   \\
\thickhline                                              & \multicolumn{4}{c}{\textbf{Number of Source Subjects}}                                                                                                                                                                                                                                                                                                                                                                \\ \cline{2-5} 
                                                                        & \multicolumn{1}{c|}{\textit {10}}                                                      & \multicolumn{1}{c|}{\textit {20}}                                                      & \multicolumn{1}{c|}{\textit{30}}                                                      & \textit{53}                                                      \\ \hline
DeepConvNet\cite{adaptivenn}                           & \multicolumn{1}{c|}{-}                                                                                 & \multicolumn{1}{c|}{-}                                                                                 & \multicolumn{1}{c|}{-}                                                                                 & \begin{tabular}[c]{@{}c@{}}84.19\\ ($\pm$9.98)\end{tabular}                            \\ \hline
DeepConvNet (Baseline)                                                   & \multicolumn{1}{c|}{\begin{tabular}[c]{@{}c@{}}72.83*\\ ($\pm$14.22)\end{tabular}}                           & \multicolumn{1}{c|}{\begin{tabular}[c]{@{}c@{}}80.65\\ ($\pm$13.02)\end{tabular}}                           & \multicolumn{1}{c|}{\begin{tabular}[c]{@{}c@{}}82.81*\\ ($\pm$12.83)\end{tabular}}                           & \begin{tabular}[c]{@{}c@{}}84.98\\ ($\pm$12.18)\end{tabular}                           \\ \hline
DeepConvNet with CPL                                                    & \multicolumn{1}{c|}{\begin{tabular}[c]{@{}c@{}}72.67*\\ ($\pm$14.04)\end{tabular}}                           & \multicolumn{1}{c|}{\begin{tabular}[c]{@{}c@{}}80.30\\ ($\pm$12.45)\end{tabular}}                           & \multicolumn{1}{c|}{\begin{tabular}[c]{@{}c@{}}82.22*\\ ($\pm$12.11)\end{tabular}}                           & \begin{tabular}[c]{@{}c@{}}84.80\\ ($\pm$11.84)\end{tabular}                           \\ \hline
Proposed Method                                                         & \multicolumn{1}{c|}{\begin{tabular}[c]{@{}c@{}}\textbf{74.17}\\ ($\pm$14.19)\end{tabular}} & \multicolumn{1}{c|}{\begin{tabular}[c]{@{}c@{}}\textbf{81.61}\\ ($\pm$12.83)\end{tabular}} & \multicolumn{1}{c|}{\begin{tabular}[c]{@{}c@{}}\textbf{84.07}\\ ($\pm$12.25)\end{tabular}} & \begin{tabular}[c]{@{}c@{}}\textbf{85.22}\\ ($\pm$12.24)\end{tabular} \\ \hline
\end{tabular}%
}
\end{table}

\subsection{Experimental Details}

We used DeepConvNet \cite{schirrmeister2017deep} without the final classification layer as the shared feature extractor. It is a popular CNN structure for EEG decoding that consists of 4 convolution-pooling blocks. Its first convolution block has temporal and spatial convolution filters.

Our baseline classifier is trained only with cross-entropy loss. Our proposed classifier is trained with distance based cross-entropy loss, prototype loss for classes along with DCE, and PL for domains. Weights of the PLs ($\beta_1$, $\beta_2$) are set to 0.001 and the weight of the domain loss ($\alpha$) is set to 0.1.  Both the style and semantic encoders are implemented using a fully connected layer. Since our input size is [62 $\times$ 1,000], the shared feature extractor outputs a feature with size 1,400. Then both the style and semantic encoders reduce the feature size to 2. All of the prototypes are initialized to zero at the beginning of the training, and only 1 prototype is kept for a class/domain. 

The loss function was optimized by the Adam algorithm with a learning rate of 0.005. 
We also used the cosine annealing learning rate scheduler. 
Batch size $b$ for training was 16. All the networks were trained for 200 epochs, and the model with the lowest loss on the validation set was selected for evaluation.

\section{RESULTS AND DISCUSSION}

We compared our results with previous subject-dependent and subject-independent methods. To observe the effect of the number of subjects we re-implemented the baseline method. In total, we experimented with three different methods in subject-independent settings. The first one is the original DeepConvNet which is trained with the traditional softmax function and the second one is the DeepConvNet which uses CPL for task classification. Lastly, our third one is the proposed framework. The comparison of results is given in Table \ref{tab:results}. Along with our experimental results, we also provide the results for DeepConvNet from another paper \cite{adaptivenn} and some subject-specific results from other papers. 

All the methods are run 5 times with 10 subjects, 3 times with 20 subjects, 2 times with 30 subjects, and 1 time with 53 subjects. All runs contain different sets of subjects but the same sets are used for each method. Our baseline achieves $72.83\%$ average accuracy on 10 subjects, $80.65\%$ on 20 subjects, $82.81\%$ on 30 subjects, and $84.98\%$ on 53 subjects. DeepConvNet with CPL achieves $70.17\%$ average accuracy on 10 subjects, $80.30\%$ on 20 subjects, $82.22\%$ on 30 subjects, and $84.80\%$ on 53 subjects. Our proposed method achieves $74.17\%$ average accuracy on 10 subjects,  $81.61\%$ on 20 subjects, $84.07\%$ on 30 subjects, and finally $85.22\%$ on 53 subjects. We analyzed the statistical significance between the methods by measuring the \textit{p}-value. The statistical values are calculated through the Friedman test after the normality test.

Our ablation study which uses CPL for only task classification seems to be the worst in all cases compared to baseline and the proposed method. However, its performance is still comparable to the baseline.

\begin{figure}[t!]
    \centering
    \includegraphics[width =0.9\linewidth,scale =0.5]{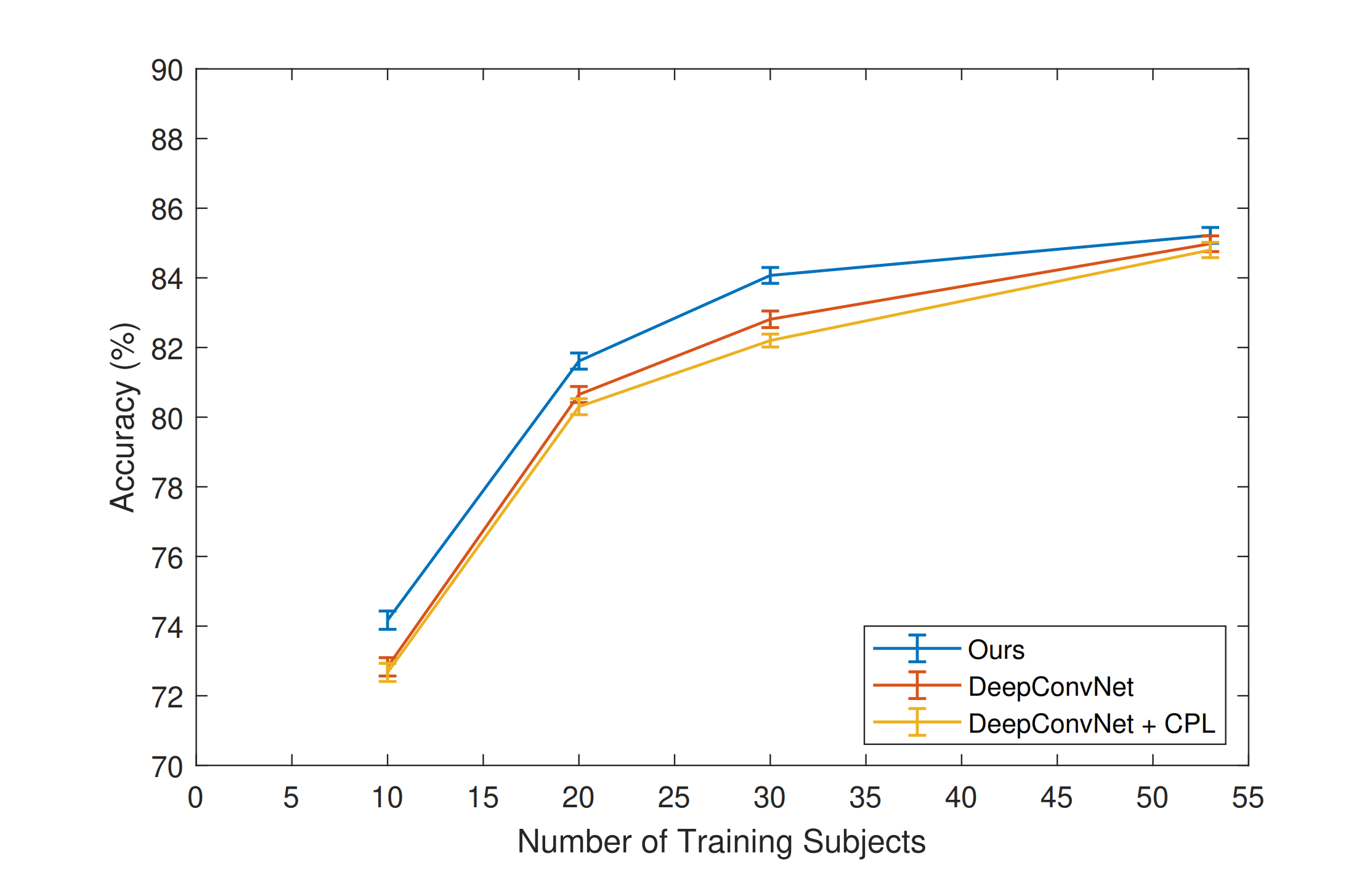}
\caption{Accuracy results across various number of training subjects. The error bars show standard errors. All the reported results are the average of different runs.}
\label{fig2:Accuracy graph}
\end{figure}

As it can be seen from the results, the proposed method outperforms both of the other methods on all sets but the gap between the methods decreases as the number of subjects increases, and it becomes almost similar when we use all the subjects. For example, the gap between the baseline and the proposed method is about $1.3\%$ in 10 subjects but it is around $0.2\%$ in 53 subjects. This change over the number of subjects can be seen better in Fig. \ref{fig2:Accuracy graph}. In our hypothesis when the number of subjects increases it gets harder for the framework to discriminate the known and unknown subjects from each other and the baseline has a better generalization performance as the amount of data increases largely.

\section{CONCLUSION}

In this paper, we proposed a prototype-based framework for EEG based BCIs that uses an open-set recognition method for unseen subjects. In our framework, the main goal is to impose cross-instance style in-variance in the same domain and reduce the open space risk on the potential unseen subject. To the best of our knowledge, the proposed framework achieved the best performance on the dataset with a subject-independent training scheme. 
In the future, we are planning to focus on adaptive schemes using the proposed framework. We will also expand our system to categorize previously unseen classes since we employ the OSR approach for both domain and class classification. In addition to this, we are planning to use the suggested framework for mental state monitoring tasks such as neurofeedback, identifying epileptic seizures, and sleep disorders.

\addtolength{\textheight}{-12cm}   




\bibliographystyle{IEEEtran}
\bibliography{references}

\begin{thebibliography}{10}
\providecommand{\url}[1]{#1}
\csname url@samestyle\endcsname
\providecommand{\newblock}{\relax}
\providecommand{\bibinfo}[2]{#2}
\providecommand{\BIBentrySTDinterwordspacing}{\spaceskip=0pt\relax}
\providecommand{\BIBentryALTinterwordstretchfactor}{4}
\providecommand{\BIBentryALTinterwordspacing}{\spaceskip=\fontdimen2\font plus
\BIBentryALTinterwordstretchfactor\fontdimen3\font minus
  \fontdimen4\font\relax}
\providecommand{\BIBforeignlanguage}[2]{{%
\expandafter\ifx\csname l@#1\endcsname\relax
\typeout{** WARNING: IEEEtran.bst: No hyphenation pattern has been}%
\typeout{** loaded for the language `#1'. Using the pattern for}%
\typeout{** the default language instead.}%
\else
\language=\csname l@#1\endcsname
\fi
#2}}
\providecommand{\BIBdecl}{\relax}
\BIBdecl

\bibitem{1}
J.~R. Wolpaw, N.~Birbaumer, D.~J. McFarland, G.~Pfurtscheller, and T.~M.
  Vaughan, ``{Brain–computer interfaces for communication and control},''
  \emph{Clin. Neurophysiol.}, vol. 113, no.~6, pp. 767 -- 791, 2002.

\bibitem{mason2007comprehensive}
S.~G. Mason, A.~Bashashati, M.~Fatourechi, K.~F. Navarro, and G.~E. Birch, ``{A
  comprehensive survey of brain interface technology designs},'' \emph{Ann.
  Biomed. Eng.}, vol.~35, no.~2, pp. 137--169, 2007.

\bibitem{lotte2010regularizing}
F.~Lotte and C.~Guan, ``{Regularizing common spatial patterns to improve BCI
  designs: unified theory and new algorithms},'' \emph{IEEE. Trans. Biomed.
  Eng.}, vol.~58, no.~2, pp. 355--362, 2010.

\bibitem{jeong2020decoding}
J.-H. Jeong, N.-S. Kwak, C.~Guan, and S.-W. Lee, ``Decoding movement-related
  cortical potentials based on subject-dependent and section-wise spectral
  filtering,'' \emph{IEEE Trans. Neural Syst. Rehabil Eng.}, vol.~28, no.~3,
  pp. 687--698, 2020.

\bibitem{discrim2016}
Y.~Wen, K.~Zhang, Z.~Li, and Y.~Qiao, ``A discriminative feature learning
  approach for deep face recognition,'' in \emph{Proc. Eur. Conf. Comput. Vis},
  2016.

\bibitem{rsvp}
D.-O. Won, H.-J. Hwang, D.-M. Kim, K.-R. Müller, and S.-W. Lee, ``Motion-based
  rapid serial visual presentation for gaze-independent brain-computer
  interfaces,'' \emph{IEEE Trans. Neural Syst. Rehabilitation Eng.}, vol.~26,
  no.~2, pp. 334--343, 2018.

\bibitem{erp}
M.-H. Lee, J.~Williamson, D.-O. Won, S.~Fazli, and S.-W. Lee, ``A high
  performance spelling system based on {EEG-EOG} signals with visual
  feedback,'' \emph{IEEE Trans. Neural Syst. Rehabilitation Eng.}, vol.~26,
  no.~7, pp. 1443--1459, 2018.

\bibitem{motorimagery}
R.~Dickstein and J.~E. Deutsch, ``{Motor Imagery in Physical Therapist
  Practice},'' \emph{Phys. Ther.}, vol.~87, no.~7, pp. 942--953, 2007.

\bibitem{robotarm}
J.-H. Jeong, K.-H. Shim, D.-J. Kim, and S.-W. Lee, ``Brain-controlled robotic
  arm system based on multi-directional {CNN}-{B}i{LSTM} network using {EEG}
  signals,'' \emph{IEEE Trans. Neural Syst. Rehabilitation Eng.}, vol.~28,
  no.~5, pp. 1226--1238, 2020.

\bibitem{multisubjectEMBC}
C.~E. Solórzano-Espíndola, E.~Zamora, and H.~Sossa, ``Multi-subject
  classification of motor imagery {EEG} signals using transfer learning in
  neural networks,'' in \emph{Conf. Proc. IEEE Eng. Med. Biol. Soc.}, 2021, pp.
  1006--1009.

\bibitem{ensembleCNN}
I.~Dolzhikova, B.~Abibullaev, R.~Sameni, and A.~Zollanvari, ``An ensemble cnn
  for subject-independent classification of motor imagery-based {EEG},'' in
  \emph{Conf. Proc. IEEE Eng. Med. Biol. Soc.}, 2021, pp. 319--324.

\bibitem{towardosp}
W.~J. Scheirer, A.~de~Rezende~Rocha, A.~Sapkota, and T.~E. Boult, ``Toward open
  set recognition,'' \emph{IEEE Trans. Pattern Anal. Mach. Intell.}, vol.~35,
  no.~7, pp. 1757--1772, 2013.

\bibitem{cpl}
H.-M. Yang, X.-Y. Zhang, F.~Yin, and C.-L. Liu, ``Robust classification with
  convolutional prototype learning,'' in \emph{2018 Proc. IEEE Comput. Soc.
  Conf. Comput. Vis. Pattern Recognit.}, 2018, pp. 3474--3482.

\bibitem{cpn}
H.-M. Yang, X.-Y. Zhang, F.~Yin, Q.~Yang, and C.-L. Liu, ``Convolutional
  prototype network for open set recognition,'' \emph{IEEE Trans. Pattern Anal.
  Mach. Intell.}, 2020.

\bibitem{openbmi}
M.-H. Lee, O.-Y. Kwon, Y.-J. Kim, H.-K. Kim, Y.-E. Lee, J.~Williamson,
  S.~Fazli, and S.-W. Lee, ``{EEG dataset and OpenBMI toolbox for three BCI
  paradigms: an investigation into BCI illiteracy},'' \emph{GigaScience},
  vol.~8, no.~5, 2019.

\bibitem{adaptivenn}
K.~Zhang, N.~Robinson, S.-W. Lee, and C.~Guan, ``Adaptive transfer learning for
  {EEG} motor imagery classification with deep convolutional neural network,''
  \emph{Neural. Netw.}, vol. 136, pp. 1--10, 2021.

\bibitem{csp}
H.~Ramoser, J.~Muller-Gerking, and G.~Pfurtscheller, ``Optimal spatial
  filtering of single trial {EEG} during imagined hand movement,'' \emph{IEEE
  Trans. Rehabil. Eng.}, vol.~8, no.~4, pp. 441--446, 2000.

\bibitem{cssp}
S.~Lemm, B.~Blankertz, G.~Curio, and K.-R. Muller, ``Spatio-spectral filters
  for improving the classification of single trial {EEG},'' \emph{IEEE Trans.
  Biomed. Eng.}, vol.~52, no.~9, pp. 1541--1548, 2005.

\bibitem{fbcsp}
K.~K. Ang, Z.~Y. Chin, H.~Zhang, and C.~Guan, ``Filter bank common spatial
  pattern ({FBCSP}) in brain-computer interface,'' in \emph{2008 IEEE Int. Jt.
  Conf. Neural Netw.}, 2008, pp. 2390--2397.

\bibitem{bssfo}
H.-I. Suk and S.-W. Lee, ``A novel bayesian framework for discriminative
  feature extraction in brain-computer interfaces,'' \emph{IEEE Trans. Pattern
  Anal. Mach. Intell.}, vol.~35, no.~2, pp. 286--299, 2013.

\bibitem{freqcnn}
O.-Y. Kwon, M.-H. Lee, C.~Guan, and S.-W. Lee, ``Subject-independent
  brain–computer interfaces based on deep convolutional neural networks,''
  \emph{IEEE Trans. Neural Netw. Learn. Syst.}, vol.~31, no.~10, pp.
  3839--3852, 2020.

\bibitem{schirrmeister2017deep}
R.~T. Schirrmeister \emph{et~al.}, ``Deep learning with convolutional neural
  networks for {EEG} decoding and visualization,'' \emph{Hum. Brain Mapp.},
  vol.~38, no.~11, pp. 5391--5420, Aug. 2017.

\end{thebibliography}
\end{document}